\documentclass[11pt]{article}
\usepackage{hyperref}
\begin{document}
\title{Scalar Transfer across a Turbulent/non-turbulent Interface in a Planar Jet}
\author{Tomoaki Watanabe$^{1, 2}$, Yasuhiko Sakai$^{1}$, Kouji Nagata$^{1}$, \\\vspace{6pt}
Osamu Terashima$^{1}$, Yasumasa Ito$^{1}$, Toshiyuki Hayase$^{3}$
\\ $^{1}$ Department of Mechanical Science and Engineering, \\\vspace{6pt}
Nagoya University, Nagoya, Japan
\\\vspace{6pt} $^{2}$ Research Fellow of the Japan Society for the Promotion of Science
\\\vspace{6pt} $^{3}$ Institute of Fluid Science, Tohoku University, Sendai, Japan}
\maketitle
\begin{abstract}
This fluid dynamics video is an entry for the Gallery of Fluid Motion of the 66th Annual Meeting of the APS-DFD. In this video, the scalar transfer across the turbulent/non-turbulent (T/NT) interface in a planar jet is investigated by using a direct numerical simulation. Visualization of the scalar flux across the T/NT interface shows that the diffusive species premixed in the ambient flow is transferred into the turbulent region mainly across the leading edge (Here, the leading edge is the T/NT interface across which the turbulent fluid turns into the non-turbulent fluid in the streamwise direction).
\end{abstract}
Free shear flows, such as wakes, jets, and mixing layers, consist of turbulent and non-turbulent regions. These two regions are divided by a sharp interface, which is called a turbulent/non-turbulent (T/NT) interface. The T/NT interface plays an important role in the development of the free shear flows\cite{westerweel2005mechanics,da2008invariants,da2010thickness,holzner2011laminar}. \par
In this study, a scalar transfer across the T/NT interface is investigated by using direct numerical simulation (DNS) of a planar jet\cite{watanabe2013visualization}. The jet Reynolds number based on the width of the jet exit ($d$) and the mean velocity at the jet exit ($U_{\mathrm{J}}$) is $Re = U_{\mathrm{J}}d/\nu=2,200$,  where $\nu$ is the kinematic viscosity. The mean streamwise velocity of the ambient flow is $U_{\mathrm{M}}=0.056U_{\mathrm{J}}$. Streamwise, cross-streamwise, and spanwise directions are represented by $x$, $y$, and $z$, respectively. We consider a diffusive species, whose instantaneous concentration is represented by ${\it{\Gamma}}$. Here, ${\it{\Gamma}}$ is defined so that ${\it{\Gamma}}=0$ in the jet exit and ${\it{\Gamma}}={\it{\Gamma}}_{0}$ in the ambient flow. Therefore, the diffusive species is entrained into the jet flow with the jet development. Governing equations for velocity $U_{i}$ and the passive scalar ${\it{\Gamma}}$ are the continuity equation, the Navier--Stokes equations, and the transport equation for ${\it{\Gamma}}$, which are written as
\begin{eqnarray}
	\frac{\partial U_{j} }{\partial x_{j}}
	&=&0,
	\label{CNEQ}\\
	\frac{\partial U_{i}}{\partial t}
	+\frac{\partial U_{j} U_{i} }{\partial x_{j}}
	&=&
	-\frac{\partial P }{\partial x_{i}}
	+\nu\frac{\partial^2 U_{i}}{\partial x_{j}\partial x_{j}},
	\label{NSEQ}\\
	\frac{\partial {\it{\Gamma}}}{\partial t}
	+\frac{\partial U_{j} {\it{\Gamma}}}{\partial x_{j}}
	&=&
	D\frac{\partial^2 {\it{\Gamma}}}{\partial x_{j}\partial x_{j}}.
	\label{STEQ}
\end{eqnarray}
Here, $P$ is the instantaneous pressure, $\nu$ is the kinematic viscosity, and $D$ is the diffusivity coefficient for ${\it{\Gamma}}$. The Schmidt number ($Sc = \nu/D$) is set to 1. These equations are solved by using a finite difference method. The size of the computational domain is $L_{x}\times L_{y}\times L_{z}=13.5\pi d\times 9.9\pi d\times 2.6\pi d$, and $N_{x}\times N_{y}\times N_{z}=700\times400\times74$ computational grid points are used. The grid is equidistant in the $x$ and $z$ directions. In the $y$ direction, the fine grid is used near the jet centerline, and the grid is stretched near the lateral boundaries. The fourth order central difference is used for the spatial discretization in the $x$ and $z$ directions, and the second order central difference is used for the spatial discretization in the $y$ direction. The continuity equation and the Navier--Stokes equations are solved by using the fractional step method. The Poisson equation is solved by the conjugate gradient method. The Crank--Nicolson method is used for the time integration of $y$ direction viscous terms, whereas the third-order Runge--Kutta method is used for the time integration of other terms. The convective boundary condition is applied to the $y$-$z$ plane at $x=L_{x}$. The free-slip boundary condition is applied to the lateral boundaries, and the periodic boundary condition is applied to the spanwise direction. Inflow velocity at the jet exit is generated by using random fluctuations. The velocity statistics which were experimentally measured at the jet exit\cite{watanabe2012simultaneous} is used to determine the boundary condition at the jet exit. The time step is set to $dt=0.02d/U_{\mathrm{J}}$. \par
According to the previous works on the T/NT interface\cite{da2008invariants,da2010thickness,bisset2002turbulent}, the T/NT interface is detected by using the vorticity norm $(\omega_{i}\omega_{i})^{1/2}$. The flow region where $(\omega_{i}\omega_{i})^{1/2}\geq 0.7 U_{\mathrm{C}}/b_{U}$ is detected as the turbulent region, whereas the flow region where $(\omega_{i}\omega_{i})^{1/2}< 0.7 U_{\mathrm{C}}/b_{U}$ is detected as the non-turbulent region. Here, $U_{\mathrm{C}}$ is the streamwise mean velocity at the jet centerline, and $b_{U}$ is the jet half-width based on the mean streamwise velocity. Therefore, the T/NT interface can be represented by the iso-surface of $(\omega_{i}\omega_{i})^{1/2}= 0.7 U_{\mathrm{C}}/b_{U}$. The same interface-detection threshhold is used by Bisset et al\cite{bisset2002turbulent}. Holzner and L{\"u}thi\cite{holzner2011laminar} derived the propagation velocity of the T/NT interface, which is written as;
\begin{eqnarray}
 v_{\mathrm{n}}=\frac{2\omega_{i}\omega_{j}S_{ij}}{|\nabla\omega^{2}|}
               +\frac{2\nu\omega_{i}\nabla^{2}\omega_{i}}{|\nabla\omega^{2}|}. 
       \label{propagation_v}
\end{eqnarray}
Here, $S_{ij}=(\partial U_{i}/\partial x_{j}+\partial U_{j}/\partial x_{i})/2$ is the component of the rate of strain tensor. In Eq. (\ref{propagation_v}), $v_{\mathrm{n}}$ is defined to be positive when the T/NT interface propagates into the non-turbulent region and negative when the T/NT interface propagates into the turbulent region. Note that $v_{\mathrm{n}}$ is the velocity of the interface movement relative to the velocity of the fluid. The scalar flux across the T/NT interface can be represented by $v_{\mathrm{n}}{\it{\Gamma}}$. The scalar flux across the T/NT interface has a great influence on the scalar entrainment process. \par
In this video, the scalar flux $v_{\mathrm{n}}{\it{\Gamma}}$ at the T/NT interface is visualized to investigate the effect of the interface orientation on the scalar transfer across the T/NT interface. Here, the T/NT interface across which the turbulent fluid turns into the non-turbulent fluid in the streamwise direction is referred to as the leading edge, whereas the T/NT interface across which the non-turbulent fluid turns into the turbulent fluid in the streamwise direction is referred to as the trailing edge. The results show that $v_{\mathrm{n}}{\it{\Gamma}}$ at the leading edge tends to be larger than that at the trailing edge. The positive value of $v_{\mathrm{n}}{\it{\Gamma}}$ represents that the diffusive species in the ambient flow is transferred into the turbulent region. Therefore, comparison of $v_{\mathrm{n}}{\it{\Gamma}}$ between the leading edge and the trailing edge shows that the diffusive species in the ambient flow is mainly transferred into the turbulent region across the leading edge. \par 

\end{document}